# Generalized kinematical symmetries of quantum phase space

V.V. Khruschov [a]

[a)] *RRC "Kurchatov Institute", Kurchatov Sq., 123182 Moscow, Russia*

Continuous symmetries generated with observables of a quantum theory in the Minkowski spacetime are discussed. An example of an originated in this way algebra of observables is the algebra of observables of the canonical quantum theory, that is contained the Lorentz group algebra and the Heisenberg algebra of phase space operators. In the general case commutation relations between observables depend on $c$, $\hbar$ and additional fundamental constants. Free field equations are considered, which are invariant with respect to generalized kinematical symmetries of the quantum phase space.

For development of a general theory of fundamental interactions it would be desired to examine in greater detail besides of properties of interactions the properties of a space-time as well [1]. Investigations along these lines have been carried out in the context of both the canonical quantum field theory [2], and various modifications of the canonical theory (e.g. papers submitted to conferences and seminars on nonlocal and nonlinear field theories and selected problems of modern theoretical physics).

There are theories with new fundamental constants other than the well known ones, $c$ and $\hbar$, among these modifications. Starting with the work [3], a theory with a fundamental length has been elaborated [4, 5]. A possible generalization of the Standard Model has been proposed in the framework of the theory with the minimal length or the maximal mass [6].

Let us consider the problem more generally, when coordinates and momenta are on equal terms and form an operator phase space. In the phase space we investigate admissible symmetries generated with observables of some quantum theory depending on extra fundamental constants other than the well known ones, $c$ and $\hbar$ [3, 7, 8]. In order to restrict a considerable list of such symmetries we require the following natural constraints [9]:

a) The generalized algebra (GA) of observables must be a Lie algebra;

b) The GA dimension must coincide with the dimension of the algebra of observables for the canonical quantum theory in the Minkowski spacetime;

c) The physical dimensions of observables, which are GA generators, should be the same as

canonical ones;

d) The GA must contain the Lorentz algebra (LA) as its subalgebra and commutation relations of the LA generators with other generators should be identical with canonical ones.

In the papers [3, 7, 8] (see also [9]) the most general algebra under the conditions a) - d) has been found and the new constants with the dimensions of length [3], mass [7] and action [8] have been introduced. The algebra of observables, which satisfy the conditions a)-d), can be presented as

$$[F_{ij}, F_{kl}] = if(g_{jk}F_{il} - g_{ik}F_{jl} + g_{il}F_{jk} - g_{jl}F_{ik}),$$
$$[F_{ij}, p_k] = if(g_{jk} p_i - g_{ik} p_j), [F_{ij}, x_k] = if(g_{jk}x_i - g_{ik}x_j),$$
$$[F_{ij}, I] = 0, [p_i, p_j] = (if/L^2) F_{ij}, [x_i, x_j] = (if/M^2) F_{ij},$$
$$[p_i, x_j] = if(g_{ij} I + F_{ij}/H), \qquad (1)$$
$$[p_i, I] = if(x_i/L^2 - p_i/H),$$
$$[x_i, I] = if(x_i/H - p_i/M^2)$$

Among relations of the system (1) Eq.1 specifies the LA, while Eqs. 2-4 specify the tensor properties for the well-known physical quantities, Eqs. 5-6 lead to the noncommutativity of p and x, Eqs. 7-9 are the generalization of the Heisenberg relation. The system of relations (1) is written in the units with c = 1 (c is the velocity of light), it contains four dimensional parameters: f, M, L, and H. But in the limiting case, when

$$M \to \infty, \ L \to \infty, \ H \to \infty,$$

the system (1) should transform to the system of relations for observables of the canonical quantum theory, so $f = \hbar$.

From mathematical point of view, the generalized algebra (1) contains, as special cases, a great number of algebras of different symmetry groups. If one evaluate the Killing – Cartan form the following condition for the algebra (1) being a semisimple algebra can be written:

$$f^2(M^2L^2 - H^2)/M^2L^2 H^2 \neq 0 \qquad (2)$$

When the condition (2) is fulfilled the GA(1) is isomorphic to a pseudoorthogonal algebra for one of the O(3,3), O(4,2), O(5,1) groups (see Table 1 below). In other cases it is isomorphic to some direct or semidirect product of a pseudoorthogonal algebra and an Abelian or an integrable algebra.

Table 1. Domains of $H^2$, $M^2$ and $L^2$ parameters corresponding to the O(2,4), O(1,5) and O(3,3) groups.

| Domains of $H^2$, $M^2$ and $L^2$ parameters | Group |
| --- | --- |
| $H^2 < M^2L^2$, $M^2 > 0$, $L^2 > 0$ | O(2,4) |
| $H^2 < M^2L^2$, $M^2 < 0$, $L^2 < 0$ | O(2,4) |
| $M^2 > 0$, $L^2 < 0$ or $M^2 < 0$, $L^2 > 0$ | O(2,4) |
| $H^2 > M^2L^2$, $M^2 > 0$, $L^2 > 0$ | O(1,5) |
| $H^2 > M^2L^2$, $M^2 < 0$, $L^2 < 0$ | O(3,3) |

For the pseudoorthogonal algebras irreducible representations are determined with the help of eigenvalues of the three Casimir operators:

$$K_1 = \varepsilon_{ijklmn} F^{ij} F^{kl} F^{mn}, \quad K_2 = F_{ij} F^{ij}, \quad K_3 = (\varepsilon_{ijklmn} F^{kl} F^{mn})^2 \qquad (3)$$

For instance, the second-order invariant operator $K_2$ in terms of I, p, x and F can be represented in the form:

$$C_2 = \Sigma_{i<j} F_{ij} F^{ij} (1/M^2 L^2 - 1/H^2) + I^2 + (x_i p^i + p_i x^i)/H - x_i x^i/L^2 - p_i p^i/M^2 \qquad (4)$$

Apart from mathematical properties which have been presented in the Refs. [9, 10] the generalized algebra (1) is the object of interest to the modern physical applications as well. For instance, in paper [11] a suggestion is made to apply the GA(1) in classical physics at the astronomical scales.

We consider possible applications of the GA (1) to quantum phenomena at microscales [12, 13]. In this case it is convenient to use the quantum constants $\kappa = \hbar/H$, $\lambda = \hbar/M$, $\mu = \hbar/L$ and to write the algebra (1) in the natural units with $c = \hbar = 1$.

$$\begin{aligned}
&[F_{ij}, F_{kl}] = i(g_{jk} F_{il} - g_{ik} F_{jl} + g_{il} F_{jk} - g_{jl} F_{ik}), \\
&[F_{ij}, p_k] = i(g_{jk} p_i - g_{ik} p_j), \quad [F_{ij}, x_k] = i(g_{jk} x_i - g_{ik} x_j), \\
&[F_{ij}, I] = 0, \quad [p_i, p_j] = i\mu^2 F_{ij}, \quad [x_i, x_j] = i\lambda^2 F_{ij}, \\
&[p_i, x_j] = i(g_{ij} I + \kappa F_{ij}), \quad [p_i, I] = i(\mu^2 x_i - \kappa p_i), \quad [x_i, I] = i(\kappa x_i - \lambda^2 p_i)
\end{aligned} \qquad (5)$$

In the general case one may classify generalized quantum fields (GQF) as the fields which form a space for irreducible representation of GA (5). For the pseudoorthogonal algebra GQF should obey the following equation among others:

$$[\Sigma_{I<j} F_{Ij} F^{Ij} (\lambda^2 \mu^2 - \kappa^2) + I^2 + \kappa(x_i p^i + p_i x^i) - \mu^2 x_i x^i - \lambda^2 p_i p^i] \Phi = 0 \qquad (6)$$

The Eq. (6) is the modification of the Klein-Gordon-Fock equation of the canonical quantum field theory.

Let us apply the GA (5) for description of color particles such as quarks or gluons. Then additional constraints should be required for the form of GA (5). On account of CP-invariance of strong interactions the constraint $\kappa = 0$ holds [12]. Moreover, the presence of a nonzero $\lambda$ value causes some inconsistencies in the quark descriptions inside hadrons and is superfluous [13]. Thus we put $\kappa = \lambda = 0$. In this case denoting $\mu$ as $\mu_s$ the following nonzero commutation relations (besides of the standard commutation relations with the Lorentz group generators) take place:

$$\begin{aligned}
&[p_i, p_j] = i\mu_s^2 F_{ij}, \\
&[p_i, x_j] = i g_{ij} I, \\
&[p_i, I] = i\mu_s^2 x_i
\end{aligned} \qquad (7)$$

From these relations it immediately follows nonzero uncertainties for results of simultaneous measurements of quark momentum components. For instance, let $\psi_{1/2}$ is a quark state with a definite value of its spin component along the third axis. Consequently, $[p_1, p_2] = i\mu_s^2/2$, thus

$$\Delta p_1 \Delta p_2 \geq \mu_s^2 / 4 \qquad (8)$$

and if $\Delta p_1 \sim \Delta p_2$, one gets $\Delta p_1 \geq \mu_s / 2$, $\Delta p_2 \geq \mu_s / 2$, i.e. the transversal quark momentum components are not measurable simultaneously.

In the framework of the quark model a rough estimation indicates that the $\mu_s$ value lies in the neighborhood of the 0.5 GeV. To find more precise number one can use a quark equation of Dirac-Gursey-Lee type [14, 15, 13].

$$[\gamma_i (p_0^i + dp_0^k L_k^i + i\mu_s \gamma^i/2) + 2i\mu_s S_{ij}(L^{ij} + S^{ij})]\psi = m\psi, \qquad (9)$$

where $p_0 + dp_0 L = p_F$ is the space-time total momentum [16], $d = \mu_s / m_0$, $p_0$ and $L$ have forms of the usual generators of translations and Lorentz transformations in the Minkowski spacetime respectively, and $p_0^2 = m_0^2$, $m_0$ is a current quark mass, $m$ is a constituent quark mass.

To estimate a value of $\mu_s$ on the basis of $m$ and $m_0$ values we use a ground quark state $\psi_0$ in a meson so the $L^{ij}\psi_0$ contribution can be neglected. By this means using the Eq. (9) one obtains the approximate relation:

$$m \cong m_0 + 2i \mu_s \qquad (10)$$

To account for the well-known inequality $m > m_0$ $\mu_s$ should be pure imaginary negative. It follows from the correspondence for ranges of parameters and pseudoorthogonal groups written above (Table 1) that the algebra under consideration is isomorphic to the algebra of the AdS group O(2,3).

Now $m$ values can be served, which have been obtained in the independent quark model (IQM) with the hadron spectroscopy data [17]. Moreover the same mass values of constituent quarks have been used for an evaluation of neutrino mixing angles consistent with experimental data [18]. Then if we pick out from the high energy physics data $m_0 \cong 2$ MeV for current $u$-quark mass and with the help of IQM $m \cong 316$ MeV we obtain $|\mu_s| \cong 157$ MeV. Thus we can evaluate mass values of the d-, s-, c- and b-current quarks on the scale ~1 GeV, which agree with the values obtained in the QCD framework [19].

In conclusion it may be noted that further investigations of Generalized Algebra (5) and properties of solutions of the equations (6) and (9) are important objectives for an achievement of a mathematical completeness of this approach as well as other physical applications.

The work is supported with the grant # 33 for fundamental research of the Kurchatov Institute in 2008 year.